\documentclass[eqsecnum,aps,showpacs]{revtex4}
\def\beq{\begin{equation}}
\def\eeq{\end{equation}}

\begin{document}

\title{
More on Tanaka-Tagoshi Parametrization of post-1PN\\ 
Spin-Free Gravitational Wave Chirps: \\
Equispaced  and Cardinal Interpolated Lattices
}
\author{R.P. Croce and Th. Demma}
\affiliation{Wavesgroup, D.I.$^{3}$E., University of Salerno, Italy}
\vspace*{-.7cm}
\author{V. Pierro and I.M. Pinto} 
\affiliation{Wavesgroup, University of Sannio at Benevento, Italy}
\date{\today}

\begin{abstract}
The conclusions obtained in  gr-qc/0101067  are shown to be valid 
also if the full 2.5PN  expansion of the chirp phase is used.
\end{abstract}


\maketitle

In a recent paper \cite{TaTa} we shew that
the  spin-free gravitational wave templates parametrization
introduced by Tanaka and Tagoshi in \cite{TaTa1} is 
effective to set up {\em uniformly spaced} post-1PN
template lattices subject to a given minimal
match constraint. 
Lattice-uniformity makes cardinal interpolation
of the match \cite{card1}, \cite{card2} 
in the $2D$ Tagoshi-Tanaka 
parameter space straightforward, yielding 
a reduction in the 2PN template density
and total number by a factor $\sim 4$ at
$\Gamma=0.97$.

In \cite{TaTa}  the  2PN  approximant
of the (spectral, spin-free) chirp  phase has been used.
On the other hand, in \cite{TaTa1}   the 2.5PN
phase (originated in  \cite{Blanchet}) has been used,
which contains (besides an  irrelevant $f$-independent term,
which is  absorbed in the unknown  coalescency phase $\phi_C$)
the additional  term: 
\beq
\frac{\pi}{128 \eta} 
\left(\frac{38645}{252} + 5 \eta\right)
\cdot \log(f/f_0),
\label{eq:logterm}
\eeq
where $\eta=m_1m_2/m^2$ is the symmetric mass-ratio, and
$f_0$ is a suitable scaling frequency \cite{irrelevant_scaling}.
It has been pointed out  \cite{Blanchet}  that  the 2.5PN phase 
agrees, in the  limit  $\eta \rightarrow 0$, 
with the one obtained by Tagoshi and Sasaki 
in the (perturbative) test mass-limit solution 
of the spin-free relativistic two-body problem \cite{test_mass}. 
While this might be a good reason to include the 2.5PN 
term  (\ref{eq:logterm}) in the template phase, 
it does {\em not} result into better overlaps with
exact (numerically computed) waveforms.
Indeed,  as shown in \cite{DamIyeSat},  2.5PN templates 
yield generally  {\em poorer} overlaps (and larger  biases) 
as compared to 2PN,
in view of the peculiar (oscillating) behaviour 
of standard  PN  approximants \cite{DamIyeSat}.

However, the conceptual question  is  posed whether the results in \cite{TaTa}  still hold if the  2.5PN  logarithmic term (\ref{eq:logterm})  is included in the template phase.
The purpose of this note is to show that  the  main conclusions in \cite{TaTa} do apply in this case too, though a number of 
differences are noted, which are summarized below.

The major difference, after the inclusion of 
the  2.5PN  logarithmic term (\ref{eq:logterm})
in the phase, is  a moderate increase in the area of the
Tanaka-Tagoshi (flattened) manifold ${\cal T}$,
yielding a correspondingly larger number of templates,
as seen by comparing Fig. 1 and Table I of this 
note to Fig. 3 and Table III of \cite{TaTa},
respectively.

The gaussian curvature of the spin-free manifold 
retains the same order of magnitude,  
as seen by comparing  Fig. 2 of this note
to Fig.s 1 and 2 of \cite{TaTa}.

The main relevant conclusions  in \cite{TaTa} do {\em not} change. 
As shown in \cite{TaTa}, the minimal match degradation 
due to using the (flattened) Tanaka-Tagoshi (spin-free) parameter-space manifold in place of the (curved) true  one, 
can be gauged in terms of the quantity  $\eta$ defined in Eq. (III.6 )
of \cite{TaTa}.  Even using  the 2.5PN  phase, 
the quantity $\eta$  is still of the same order of magnitude
as the corresponding one computed in \cite{TaTa}.
This is seen by comparing Fig. 3 of this 
note to Fig.s  5 and 6  of \cite{TaTa}.

\section*{Acknowledgements}
The Authors thank dr. A. Pai  (IUCAA) for having posed the problem, 
and dr.  B. Owen (Univ. of Wisconsin Milwaukee) for pinpointing an error in the first version of this note.


$$~$$
$$~$$
$$~$$
\begin{center}
\begin{tabular}{|c|c|c|}
\hline\hline
Antenna &  Simplex area $[sec^2]$ & No. of templates at  $\Gamma=0.97$\\
\hline
TAMA300  &  $10266$   & $1.71\cdot 10^5$\\                                                                                                                                                                                                                                                                                                                                           
\hline
GEO600  & $57093$ & $9.51\cdot 10^5$\\
\hline
LIGO-I  & $24732$  & $4.12\cdot10^5$\\
\hline
VIRGO  &   $776392$ &  $1.29\cdot 10^7$\\                                                                                                                                                   
\hline\hline
\end{tabular}
$$~$$
Table I - 2.5PN flat simplex area and  number of  templates at $\Gamma=.97$,\\ 
for $0.2M_{\odot} \leq m_1 \leq m_2 \leq 10M_{\odot}$.  
\end{center}


\begin{thebibliography}{99}

\bibitem{TaTa}{R.P. Croce, Th. Demma, V. Pierro and I.M. Pinto, gr-qc/0101067; in print on Phys. Rev. {\bf D63}  (2001).}
\bibitem{TaTa1}{T. Tanaka and H. Tagoshi, Phys. Rev. {\bf D62}, 082001 (2000).}
\bibitem{card1}{R.P. Croce, Th. Demma, V. Pierro and I.M. Pinto, Phys. Rev. {\bf D62}, 124020 (2000).}
\bibitem{card2}{R.P. Croce, Th. Demma, V. Pierro, I.M. Pinto, D. Churches and B.S. Sathyaprakash, Phys. Rev. {\bf D62}, 121101(R) (2000).}
\bibitem{Blanchet}{L. Blanchet, T. Damour, B. Iyer, C.M. Will and A. Wiseman, Phys. Rev. Lett., {\bf 74}, 3515 (1995). }
\bibitem{irrelevant_scaling}{Scaling $f$ to $f_0$ in the argument of 
(\ref{eq:logterm}) entails an irrelevant  constant additive phase 
which is  absorbed in the unknown  coalescency phase $\phi_C$.}
\bibitem{test_mass}{H. Tagoshi and M. Sasaki, Progr. Th. Phys., {\bf 92}, 745 (1994).}
\bibitem{DamIyeSat}{T. Damour, B. Iyer and B.S. Sathyaprakash, Phys. Rev. {\bf D57}, 885 (1998).}
\end{thebibliography}
\end{document}